\documentclass[preprint2]{aastex63}

\received{}
\revised{}
\accepted{}
\submitjournal{}

\shorttitle{}
\shortauthors{}

\usepackage{float}

\begin{document}

\title{An ALMA search for high albedo objects among the mid-sized
Jupiter Trojan population}

\author{Anna M. Simpson}
\affil{Department of Astronomy and Department of Physics \\ University of Michigan \\ Ann Arbor, MI 48109, USA}

\author[0000-0002-8255-0545]{Michael E. Brown}
\affiliation{Division of Geological and Planetary Sciences\\
California Institute of Technology\\
Pasadena, CA 91125, USA}

\author{Madeline J. Schemel}
\affiliation{Division of Geological and Planetary Sciences\\
California Institute of Technology\\
Pasadena, CA 91125, USA}

\author{Bryan J. Butler}
\affil{National Radio Astronomy Observatory, Socorro NM 87801 (U.S.A.)}

\correspondingauthor{M.E. Brown}
\email{mbrown@caltech.edu}

\begin{abstract}
We use ALMA measurements of 870 $\mu$m thermal emission from a sample
of mid-sized (15-40 km diameter) 
Jupiter Trojan asteroids to search
for high albedo objects in this population.
We calculate the diameters and albedos of
each object using a thermal model which also
incorporates {contemporaneous} Zwicky Transient Facility 
photometry to accurately measure the absolute magnitude at the time
of the ALMA observation. We find that while many albedos
are lower than reported from WISE, several small Trojans
have high albedos independently measured both from ALMA and 
from WISE. The { number} of these high albedo objects is
approximately consistent with expectations of the number
of objects { that} recently have undergone large-scale impacts,
suggesting that the interiors of freshly-crated Jupiter Trojans could contain high albedo materials such as ices.

\end{abstract}


\section{Introduction} \label{sec:intro}
The Jupiter Trojan asteroids, at the intersection of the inner and outer solar
system, hold some of the keys to understanding the formation and
early dynamical evolution of the entire solar system. In the modern incarnation
of the Nice model of dynamical instability, the Jupiter Trojans formed
beyond the region of the giant planets, 
at the same location as the objects currently in the Kuiper belt
\citep{2005Natur.435..462M, 2010CRPhy..11..651M}.
Previous hypotheses had instead suggested that the Jupiter Trojans formed 
in the main asteroid belt or closer to the Jupiter system, and that they
share no relationship with objects in the Kuiper belt \citep{2002Icar..159..328M}.
Connecting the compositions of the
Jupiter Trojans and the Kuiper belt objects (or finding that they are
not connected) is critical to answering fundamental questions about
the nature of the early dynamical evolution of the solar system.

Unfortunately, we have essentially zero knowledge of the interior compositions of Jupiter Trojans.
After 4 billion years of space weathering, the surfaces of the 
Jupiter Trojans are now covered with a thick
irradiated mantle that mostly
defies spectroscopic identification 
\citep{2007Icar..190..622F, 2011AJ....141...25E,2016AJ....152..159B}. 
One solution to the lack of access to the interiors of Jupiter Trojans is to search 
for objects that have suffered recent massive collisions and have some of their
interior materials freshly exposed. 

Freshly exposed interior material from the main asteroid belt and from the Kuiper belt have very different compositions. The Haumea 
collision family -- the one known collisional family in the Kuiper belt -- is composed
of objects with nearly pure water ice surfaces \citep{2007Natur.446..294B}, even
billions of years after the family-forming impact \citep{2007AJ....134.2160R}. 
These objects are unique in the observed Kuiper belt. In contrast, even the youngest
known collisional family in the main belt, the Karin family, at only $\sim$6 Myr old, has family members 
whose surfaces are indistinguishable from the background population
\citep{2009Icar..199...86H}. We expect this general
principle to hold: fresh collisional fragments of objects from the inner and outer
solar system should be distinguishable by their distinct compositions.

If the interiors of Jupiter Trojans are composed of outer
solar system ices, we should expect that fragments left over after a
catastrophic collision would have high albedos, yet no evidence 
exists for albedo differences between members
of the Trojan asteroids’ best-known collisional family –- the Eurybates
family \citep{2010Icar..209..586D} –- and the rest of the Trojan
population. Either the interiors of these objects do not contain high-albedo material, or irradiation, devolatilization, and
space weathering
have hidden the albedo signature of the fresh materials over time. The
Eurybates family is presumably ancient, but impacts younger than the
$\sim$100 Myr timescale of space weathering \citep{1987JGR....9214933T,
2006ApJ...644..646B} could still have regions of elevated albedo.

Collisonal models suggest that collisional fragments from $\sim$100 Myr
old catastrophic impacts
are likely common among the smallest
known Jupiter Trojans ($\sim$1 km) \citep{2007A&A...475..375D}, 
but these objects are prohibitively
faint for detailed study.
While larger (and thus brighter) collisional fragments are  
rare, a small number of them must exist in the Trojan population.
Based
on collisional models, about 5\% of Trojans in the 20--30 km size range will have had catastrophic impacts in the past 100 Myr
\citep{2007A&A...475..375D}, with many more having significant
sub-catastrophic cratering events. It is in this size range where, if
the interiors of Jupiter Trojans contain high albedo materials such as
ices, we might expect to find a small number of objects with elevated
albedos.
If such larger recent collisional fragments
can be found, their larger size and brightness
would make them attractive targets for 
detailed spectroscopy to understand the interior compositions
of Jupiter Trojans.

\begin{figure}
\epsscale{1.4}
\plotone{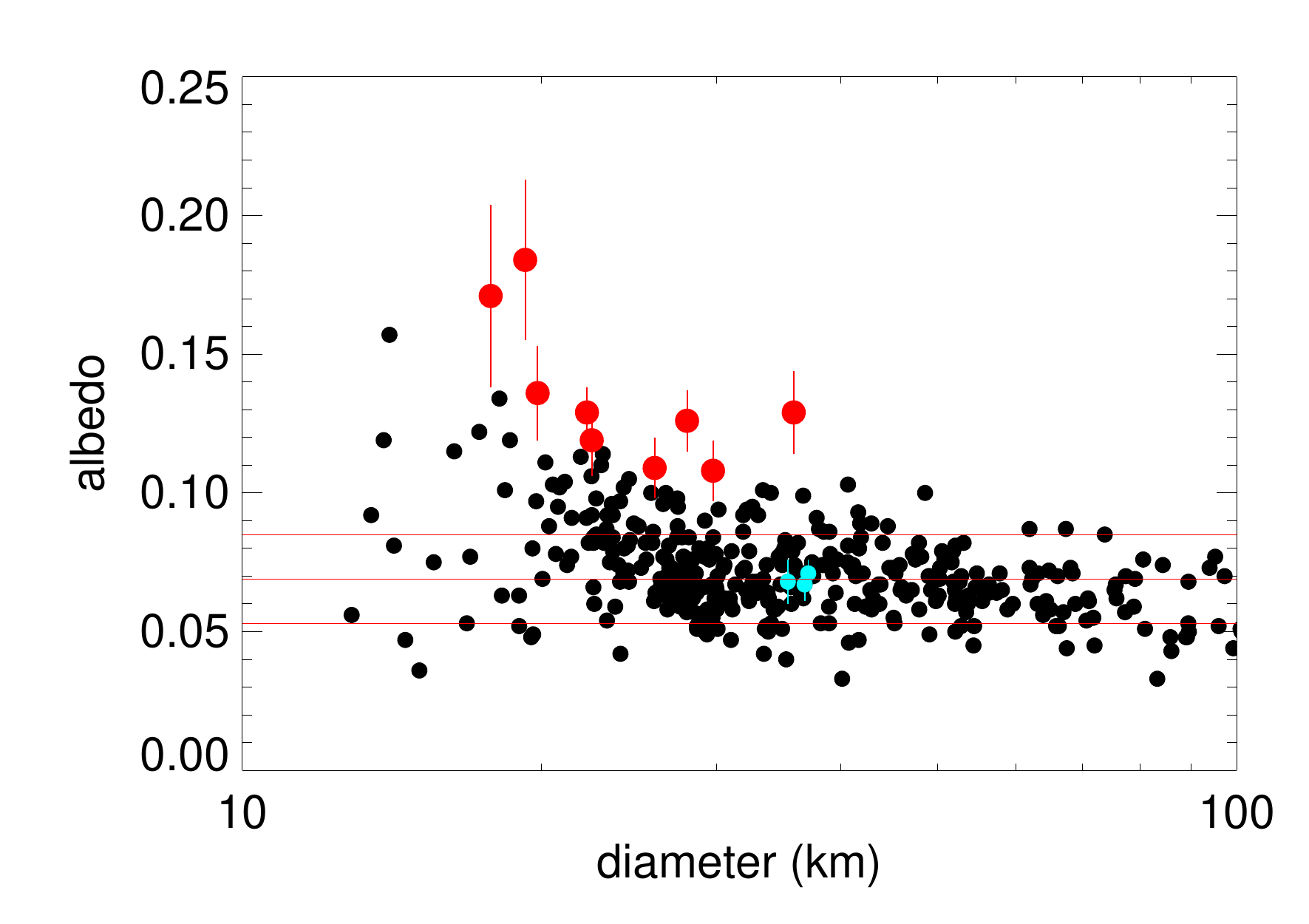}
    \caption{{ Derived albedos and diameters based on WISE observations \citep{2012ApJ...759...49G}. The black
    points show all Trojans for which 
    the WISE-derived albedo uncertainties are 0.02 or lower. 
    The median and 1$\sigma$ variation of albedo of these
    Trojans are shown as red horizontal lines. The
    nine objects shown in red all have WISE-derived albedos and 
    uncertainties that appear to be 3$\sigma$ or more above 
    the median Trojan albedo.
    Also shown, in cyan, are three objects measured
    by WISE to have typical albedos;
    these targets are used as emissivity calibrators
    for the survey.}}
    
    \label{WISE}
\end{figure}

At first glance, measurements from the WISE spacecraft appear to support the
expectation that a small number of 20--30 km Jupiter Trojans indeed
have 
elevated albedos, but the reliability of { the derived albedos} is uncertain. 
WISE detected 476 Jupiter Trojan asteroids at some
combination of wavelengths including 3.3, 4.6, 12.1, and 22.2 $\mu$m
\citep{2012ApJ...759...49G}. The two longest wavelengths sample the short-wavelength tail of the blackbody emission of these objects, allowing
radiometric determination of their diameters and albedos. 
Unlike in the main asteroid belt or even in the Kuiper belt, 
the albedos of the Trojans appear strikingly uniform (at least among the
well-sampled large objects), with a median albedo of 0.069.  
There is a hint in the WISE data that albedos may rise below a diameter of 30 km, 
but much of this apparent rise { could be} driven by the larger
uncertainties in the thermal fluxes of these smaller objects, { the
positive bias in albedo, and bias from the optically-selected sample}. In addition, measuring the albedo
using only the
exponentially changing short-wavelength end of the 
blackbody emission is highly sensitive
to model assumptions and parameters. 
With these caveats, a small number of objects appear to have 
sufficiently high albedos and sufficiently small uncertainties that
their albedos are statistically inconsistent with the median value (Figure \ref{WISE}). 
Visible and infrared spectroscopy of a subset of these from the Very Large Telescope
shows nothing unusual, but the limits on detection of water ice, 
for example, which
has only relatively weak absorption features shortward of 2.5 $\mu$m,  are not strong \citep{Marsset_2014}.

\begin{deluxetable*}{lcccccc}
\label{tab:obsinfo}
\tablecaption{Observational Details for each object. The observation time and length is given, as well as the heliocentric and geocentric distances of the object alongside the flux density for each observation.}
\tablehead{\colhead{Object} & \colhead{Midpoint} & \colhead{Total time} & \colhead{Geocentric} & \colhead{Heliocentric} & \colhead{Phase Angle} & \colhead{Flux Density}\\
\colhead{} & \colhead{} & \colhead{} & \colhead{distance (AU)} & \colhead{distance (AU)} & \colhead{(degrees)} & \colhead{(mJy)}}
\startdata
05123 (1989 BL)  &2019  Oct 06 23:02 & 5 min & 5.08 & 5.66& 8.7 & 1.39 $\pm$ 0.07\\
08125 (Tyndareus)&2019 Oct 06 02:52 & 5 min & 4.01 & 4.96 & 4.0 & 1.4 $\pm$ 1.0\\
                 &2019 Oct 07 02:45 & 5 min & 4.01 & 4.96 & 4.1 & 1.18 $\pm$ 0.07 \\
11488 (1988 RM11)&2019 Oct 08 20:28 & 18 min & 6.21 & 5.33 & 4.6 & 0.24 $\pm$ 0.06 \\
                 &2019 Oct 09 15:48 & 18 min & 6.22 & 5.33 & 4.5& 0.18 $\pm$ 0.04\\
13331 (1998 SU52)&2019 Oct 06 23:32 & 24 min & 4.62 & 5.21 & 9.4 & 0.22 $\pm$ 0.03\\
13372 (1998 VU6) &2019 Oct 06 02:28 & 18 min & 4.36 & 5.12 & 7.9 & 0.51 $\pm$ 0.05\\
13694 (1997 WW7) &2019 Oct 06 03:11 & 12 min & 4.04 & 5.01 & 2.7 & 0.86 $\pm$ 0.06 \\ 
18054 (1999 SW7) &2019 Oct 08 20:03 & 5 min & 5.98 & 5.15 & 5.7& 0.50 $\pm$ 0.09 \\
                 &2019 Oct 10 18:36 & 5 min & 5.99 & 5.15 & 5.5& 0.68 $\pm$ 0.09 \\
18137 (2000 OU30)&2019 Oct 06 15:18 & 5 min & 6.15 & 5.15 & 0.7 & 0.43 $\pm$ 0.09 \\
                  &2019 Oct 09 14:53 & 5 min & 6.15 & 5.15 & 0.9 & 0.47 $\pm$ 0.07 \\
18263 (Anchialos) &2019 Oct 06 04:03 & 18 min & 4.24 & 5.20 & 3.4 & 0.34 $\pm$ 0.05 \\
                  &2019 Oct 07 03:19 & 18 min & 4.25 & 5.20 & 3.6  & 0.33 $\pm$ 0.04 \\
24452 (2000 QU167)&2019 Oct 10 16:01 & 18 min & 6.34 & 5.43 & 4.1 & 0.15 $\pm$ 0.05 \\
                  &2019 Oct 10 17:57 & 18 min & 6.34 & 5.43 & 4.1 & 0.21 $\pm$ 0.05 \\
32501 (2000 YV135)&2019 Oct 10 15:36 & 5 min & 5.79 & 4.87 & 4.1 & 0.5 $\pm$ 0.2 \\
                  &2019 Oct 10 17:02 & 5 min & 5.80 & 4.87 & 4.0 & 0.6 $\pm$ 0.1 \\
42168 (2001 CT13) &2019 Oct 06 03:34 & 12 min & 4.05 & 4.98 & 4.6 & 0.31 $\pm$ 0.05\\
\enddata
\end{deluxetable*}

These high-albedo objects could be the collisional fragments of recent impacts. 
Indeed, models suggest that this is approximately the expected number of
objects that should have had catastrophic impacts in the past 100 Myr.
Unfortunately, an alternative possibility is that the putative high-albedo objects are normal low albedo Trojans and that observational and modeling uncertainties have been underestimated. Determining sizes and albedos of objects is particularly hard when all thermal observations are on the steep Wein section of the blackbody curve where small changes in modeling assumptions can lead to exponential changes in flux density.
Calibration of the WISE flux densities to better than 10\% is hampered
by uncertainties in the color correction
required for low temperatures of these distant targets
\citep{2010AJ....140.1868W}.

In order to independently estimate the albedos of these potentially
high-albedo objects, we obtained 870 $\mu$m radiometry of 9
of these objects using the Atacama Large Millimeter Array (ALMA). In addition, we observe 3 Trojan asteroids with 
well-measured albedos consistent with the albedos
measured for the larger members of the population and use
these measurements to calibrate the millimeter emissivity.
Observing these targets from ALMA has two main advantages: first, we can 
control the signal-to-noise (S/N) of the detections by varying 
the exposure time, rather than having to rely on the uniform WISE survey, 
and thus are not as affected by a falling S/N at smaller sizes. 
Second, at these wavelengths, which cover the Rayleigh-Jeans portion of 
the blackbody, the flux density is only linearly, 
rather than exponentially, sensitive to surface temperature, 
making the observations significantly less sensitive to modeling assumptions. 
We use the observations as well as contemporaneous
observations of the visible flux from these objects 
using the Zwicky Transient Facility (ZTF) archive to
investigate the albedos of these potentially high albedo 
objects.

\section{OBSERVATIONS AND DATA REDUCTION}

The observations of the Trojan asteroids were taken with the main array
of ALMA, which is composed of up to
50 12-meter diameter antennas spread across the Altiplano in the high
northern Chilean Andes.  ALMA can operate in 7 frequency windows, from
$\sim$90 to $\sim$950 GHz. The observations presented here were taken in
Band 7, near 340 GHz, in the ``continuum'' (or ``TDM'') mode, with
standard frequency tuning.  This set up results in four spectral windows with
frequencies of 335.5--337.5 GHz, 337.4--339.4 GHz,
347.5--349.5 GHz, and 349.5--351.5 GHz. In the final steps of the data
reduction, we averaged over the entire frequency range, and use 343.5
GHz (872.8 $\mu$m) as the observation frequency in our thermal modelling.

We observed with ALMA on dates from October 6 to 10 of 2019
(Table \ref{tab:obsinfo}).  For these
observations, there were between 41 and 46 antennas, mostly in the C43-4
configuration.  In the earlier observations (October 6-7), some antennas
were on pads isolated from the rest of the array, and,
to be conservative, we exclude those antennas
from later analysis, thus
between 38 and 46 antennas are used in the final
analysis.  The
C43-4 configuration has a maximum antenna spacing of $\sim$784 m, giving a maximum resolution on the sky of $\sim$300 mas at our observing
frequency, much larger than the $\sim$ 20 mas apparent diameter of any of the Trojan asteroids.

Observations lasted from 16 to 40 minutes in duration, including all
calibration overheads, which resulted in 5 to 24 minutes on the
asteroid.  One of six ``grid calibrators'' was used as the
absolute flux density scale calibrator for the observations, depending
on where in the sky the targets were at the time.  These grid calibrators are
regularly monitored against the main flux density scale calibrators for
ALMA.  The absolute flux density scale calibration which results is
believed to be good to 5\% in Band 7 for ALMA, but can in some cases be
worse \citep{2020_Francis}.  Nearby point-like calibrators were used to
calibrate the phase of the atmosphere and antennas as a function of
time.

Initial calibration of the data was provided by the ALMA observatory,
completed in the CASA reduction package via the ALMA pipeline
\citep{2014_Muders}. 
We exported the provided visibilities from CASA
and continued the data reduction in the AIPS reduction package.  We
flagged outlier antennas, then made a decorrelation correction to each
observation.  Such a correction is necessary because even after normal calibration,
phase errors remain in the data, and an estimate of their effect on the
visibilities must be made in order to derive meaningful flux densities
\citep{2017_Thompson}.  If the asteroids had enough flux density to
self-calibrate, we could use that to correct these phase errors, but
they do not.  Fortunately, however, ALMA included  ``check
sources" in each of these observations.  These are nearby point sources
that do have enough flux density to be self-calibrated, and are
observed and calibrated in the same way as the target source.  So
estimating flux density before self-calibration, then after, for these
check sources gives a correction factor which we applied to the data.
We then did a fit of the asteroid visibilities to a point source, along
with making an image.  The fit to the visibilities provides a more
reliable flux density, since it is not subject to imaging deconvolution
errors.  But in all cases, the fit to the visibilities and the fit to a
Gaussian in the image agreed to within one-sigma.  For the asteroids
which had two separate observations, we fitted each one separately, then
combined the data and fit that as well.  The resultant fitted flux
densities are shown in Table \ref{tab:fluxandmag}.

\begin{deluxetable*}{lccc}
\label{tab:fluxandmag}
\tablecaption{Rotation periods, combined flux densities, and
absolute magnitudes for each object. 
Where no period was able to be determined using the ZTF data, the period column is left blank. If multiple periods are possible
all are listed as well as the phased
absolute magnitude for each potential period. }
\tablehead{
\colhead{} & \colhead{Period} & \colhead{Flux Density} & \colhead{Absolute} \\
\colhead{Object}&\colhead{(h)}&\colhead{(mJy)}&\colhead{Magnitude}}
\startdata
05123 (1989 BL)    & $9.897 \pm 0.008$  & $1.39\pm0.07$   & $10.1\pm 0.1$  \\
08125 (Tyndareus)  & $51.2\pm 0.2$    & $1.26  \pm 0.06$   & $11.02\pm 0.09$ \\
11488 (1988 RM11)  & ||                 & $0.20 \pm 0.03$    & $11.44 \pm 0.04$ \\
13331 (1998 SU52)  & $373.5 \pm 13.8$ & $0.22 \pm 0.03$  & $11.59 \pm 0.06$ \\
13372 (1998 VU6)   & ||                 & $0.51 \pm  0.05$ & $11.34 \pm 0.05$  \\
13694 (1997 WW7)   & $19.95 \pm 0.02$  & $0.86 \pm 0.06$  & $11.12 \pm 0.06$ \\
                   & $34.14 \pm 0.06$  & \texttt{"}        & $11.09 \pm 0.06$ \\
18054 (1999 SW7)\tablenotemark{a}   & ||                 & $0.59\pm 0.06$   & $10.75 \pm 0.07$ \\
18137 (2000 OU30)\tablenotemark{a}  & $12.102 \pm 0.007$ & $0.45 \pm 0.06$  & $11.26\pm 0.07 $  \\
                   & $16.20 \pm 0.01$  &\texttt{"}         & $11.27 \pm 0.09$ \\
                   & $24.47 \pm 0.05$  &\texttt{"}         & $11.17 \pm 0.07$  \\
18263 (Anchialos)  & $10.330\pm 0.006$  & $0.33\pm 0.03$   & $11.58 \pm 0.07$  \\
24452 (2000 QU167) & ||                 & $0.18 \pm 0.03$  & $11.94 \pm 0.07$ \\
32501 (2000 YV135)\tablenotemark{a}     & ||& $0.59 \pm 0.09$  & $11.38 \pm 0.07$  \\
42168 (2001 CT13)  & $4.499\pm 0.002$   & $0.31 \pm 0.05$  & $11.53\pm 0.09$  \\
                   & $4.966\pm 0.002$   & \texttt{"}        & $11.43\pm 0.07$  \\
                   & $5.540\pm 0.002$   & \texttt{"}        & $11.38\pm 0.07$ \\
                   & $6.265\pm 0.003$   & \texttt{"}        & $11.41\pm 0.07$  \\
                   & $7.206\pm 0.004$   & \texttt{"}        & $11.55\pm 0.07$  \\
                   & $8.482\pm 0.005$   & \texttt{"}        & $11.66\pm 0.07$  \\
                   & $10.308\pm 0.008$& \texttt{"}       & $11.68 \pm 0.07$  \\
                   & $13.14\pm 0.01$   & \texttt{"}        & $11.55 \pm 0.07$ \\
\enddata
\tablenotetext{a}{emissivity calibrator}
\end{deluxetable*}

\section{Zwicky Transient Facility photometry}\label{sec:dataZTF}

Radiometric measurement of diameter and albedo requires not just thermal
emission, but an accurate measurement of the visible magnitude at
the time of the thermal observation. In practice determining such
a magnitude requires determining both the rotational amplitude of the 
object and the phase at the time of the ALMA observation. These
parameters are generally unknown for our targets. Fortunately, 
ZTF was engaged in a large scale photometric
sky survey at the time of our ALMA observations. All targets were
observed a minimum of 27 times during the 2019 opposition season
with one- to several-day spacings. 
\citet{schemel_zwicky_2020} analyzed the photometric data from $\sim$1000
Trojans asteroids and developed a powerful method to extract
absolute magnitudes, colors, phase curves, and rotational amplitudes from
these sparsely sampled data. For most objects, however, they did not
attempt to determine the rotation period, which we need to correctly
phase the data. We further analyze the ZTF data
with the goal to directly determine the absolute magnitude at the time of
the observation. 

For each of the ALMA targets, we take the ZTF photometric
data from \citet{schemel_zwicky_2020},
convert it to a V-band absolute magnitude using the 
derived color and phase curve and the appropriate conversion from \citet{jester_sloan_2005} for solar colors, and subtract the average magnitude, yielding
observations that should only be affected by the light curve of the asteroid.
The time series of these residual magnitudes was then examined for a periodic
behaviour. 
For each object, the Astropy implementation of a Lomb-Scargle periodogram was
used to construct a periodogram showing the power at different frequencies. 
In many cases, we find a single strong peak with aliases at frequencies 
corresponding to frequencies separated by $\sim$1 day$^{-1}$, as expected
from these nightly-sampled data. To calibrate the significance of the
periodogram peaks, the data were shuffled 
by randomly assigning each observed photometric point to an observed observation
time. A periodgram of these shuffled data will show the effects of the noise
and the observing cadence on the periodogram. 
We perform 100 iterations of this shuffling and retain
the maximum periodogram power over the entire frequency range
over all iterations
as our limit
of significance.

{\ As an additional test of the significance of the
periodic fits, we calculate the Akaike Information
Criterion (specifically the version corrected
for smaller sample sizes, knowns as AIC$_C$).
The AIC$_C$ is defined as:

$${\rm AIC}_C \equiv 2k-2 \ln \mathcal{L_{\rm max}}+{{2k^2+2k}\over{n-k-1}}$$
where $\mathcal{L}_{\rm max}$ is the value of the maximum likelihood, 
$k$ is the number of free parameters, and $n$ is the number of data points. The model with the lower value of AIC$_C$ is prefered
by a factor of $\exp[(\Delta{\rm AIC}_C)/2],$ where $\Delta$AIC$_C$ is the difference
in AIC$_C$ between the two models \citep{Burnham_Anderson2004}, which, in our case,
is the difference between a model of constant 
magnitude and a simple sinusoidal maximum likelihood
fit at the 
selected period. In all cases where we list a
potential period fit, $\Delta$AIC$_C>$16, implying
that the periodic fit is preferred by more than a factor of 3000. In the majority of cases
$\Delta$AIC$_C>100$.}

\begin{figure*}
    \centering
    \includegraphics[width=15cm]{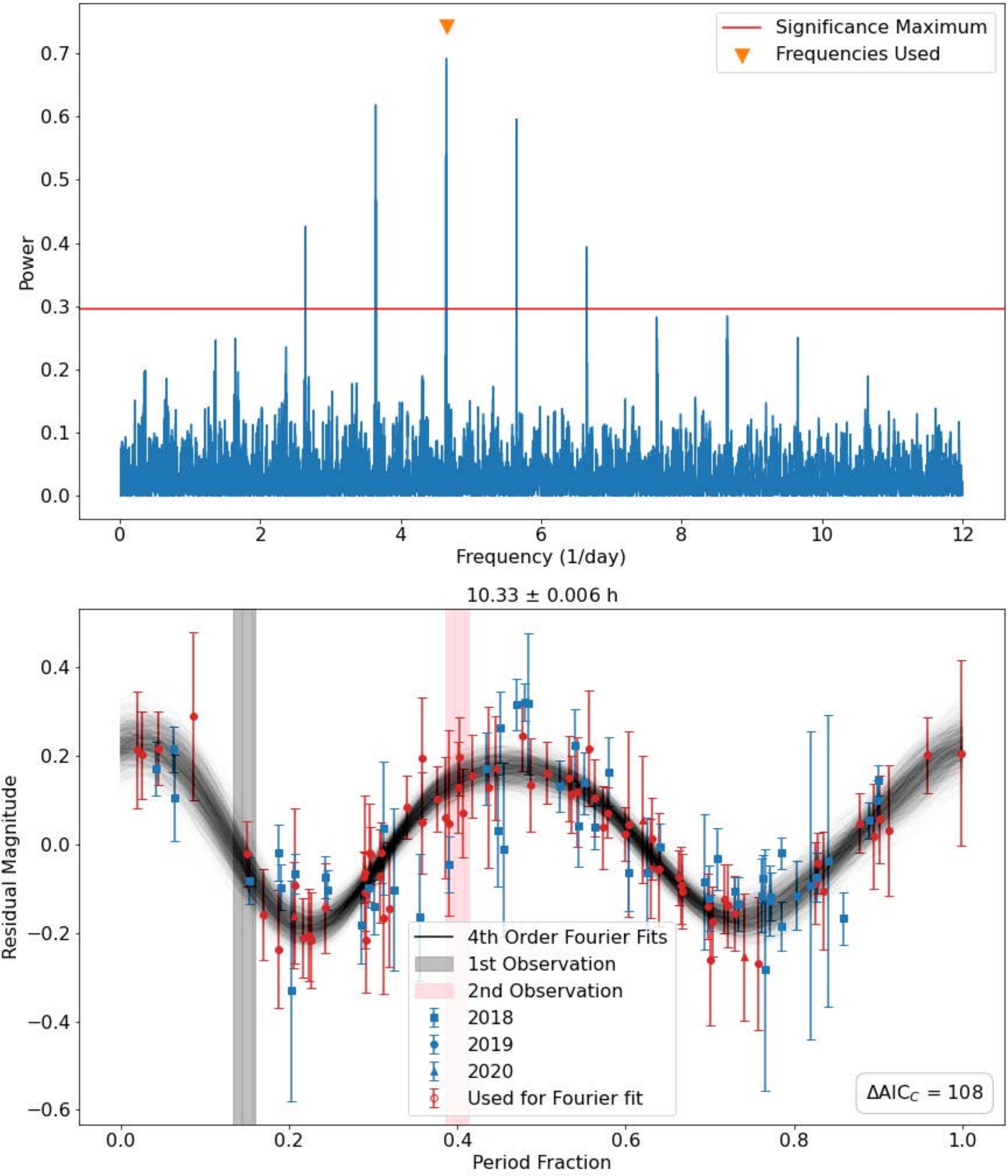}
    \caption{Periodogram and phased ZTF lightcurve for object 18263. While multiple periods are above the significance threshold, only one (the highest peak) is consistent with previous data. The ZTF data phased to that period is plotted below it. The shape of each data point indicates the year in which that data point was taken. The red points indicate which data were used to find the best fit fourth order Fourier series using an MCMC. {\ The value of $\Delta$AOI$_C$ calculated for the 
    periodic fit compared to a constant light curve is shown in the bottom right (see text for details).} Random samples of the results of this fitting process are plotted as dark lines over the data. The grey and red shaded regions indicate where in the curve the ALMA observations occurred.}
    \label{fig:lightcurves}
\end{figure*}

Fig \ref{fig:lightcurves} shows an example
of this procedure from object 18263. The other Trojan asteroids for which potential periods were found
are shown in the Appendix.

For each possible period over the significance threshold we performed
multiple steps to determine the viability of the potential period.
We first phased all data to each potential period. As expected
from the periodograms, all data were consistent with each significant period,
with the quality of the fit decaying as the periodogram power dropped.
Next, we attempted to confirm
rotational periods from published values.
Objects 5123 and 13331 were observed in the Kepler K2 mission and analyzed in
\citet{szabo_heart_2017} and \citet{ryan_trojan_2017}. 
Based on these results, all but one potential period 
could be ruled out for these two objects.
In addition, previous data for 13694, 18263, and 18137 as presented in \citet{french_photometry_2012}, \citet{french_rotation_2015}, and \citet{stephens_dispatches_2015} respectively, were used to check the viability of periods found using ZTF data. 
Previously observed data for each of these objects was obtained from the Asteroid Lightcurve Photometry Database\footnote{\url{http://alcdef.org/}} \citep{stephens_proposed_2010, warner_save_2011} 
and these magnitudes were phased to each potential period indicated by the periodogram. In cases where the additional data did not produce a consistent lightcurve when phased to a given period, that period could be discarded, thus reducing the number of potential periods for these three objects. For object 32501, many periods were above the significance threshold, and its relatively flat lightcurve prevented any assessment of whether these periods indicated actual periodic variation in the lightcurve. In addition, for object 8125, all but one potential period was discarded as the phased lightcurve was not consistent with periodic behavior.
The period for each object, or, in some cases, 
the multiple possible periods, are shown in Table \ref{tab:fluxandmag}.
Note that we assume that all light curves are mostly
due to shape and thus have two full cycle per rotation.
The listed periods are thus twice that of the 
best single-cycle periods found in the periodograms. The uncertainties on the periods listed in Table \ref{tab:fluxandmag} are better than the literature values for 5123 \citep{ryan_trojan_2017}, 18263 \citep{french_rotation_2015}, and that for the 16.2 day period for 18137 \citep{stephens_dispatches_2015}.

With periods or groups of possible periods for all objects
now determined, we determined the shape of each light
curve by using each possible period
to phase all ZTF data taken within 
6 months of the the ALMA observation
(the full
ZTF data set was not used owing to slow viewing-angle
changes over time that can change the shape
 -- but not period -- of the light curve).
We use a fourth-order Fourier series to
better fit the full asymmetric light curve from each data set.
To understand the full range of uncertainties
from such a fit we create a likelihood
model with the four Fourier parameters
as our model parameters and use 
a Markov-Chain Monte Carlo (MCMC)
procedure -- as implemented in the \texttt{emcee} package \citep{foreman-mackey_emcee_2013} --
to sample the four-dimensional phase space
of Fourier coefficients.
All MCMC chains were run to a length of 5600 samples,
more than
50 times the autocorrelation length, and the first
800 samples of each chain were discarded as burn-in.
From each Markov Chain sample we create
a light curve and determine the absolute magnitude
at the time of the observation.
The best-fit absolute magnitude 
is then taken to be the median of the 
value predicted for that time given the
sample coefficients. The uncertainties appear
essentially Gaussian thus we take the $1 \sigma$
variation in magnitude to be the 
16th and 84th percentile values of the absolute 
magnitude. 

Figure \ref{fig:lightcurves} illustrates 
this process, showing the initial 
periodogram obtained for object 18263. 
All but one potential period was discarded 
by cross-checking the periods with data in
the literature, so only one frequency is indicated 
on the periodogram. 
Beneath, we plot the phased ZTF data as
well as a random selection of MCMC light curve samples
to illustrate the uncertainties. 

Some objects have little constraint on the rotation
period and have no value listed in Table \ref{tab:fluxandmag}, 
owing either to a small light curve or
to other unknown factors, so we are able to obtain
neither a period nor a phase for the ALMA observation.
For these objects we simply take the mean absolute
magnitude for the 2019 season and use the full amplitude of
the light curve from \citet{schemel_zwicky_2020}
as the uncertainty.

\section{Thermal Model}\label{sec:TherMod}
In order to obtain values of physical properties 
-- including albedo --
based on the thermal emission observed, a thermal 
model similar to that given in \citet{harris_thermal_1998} is used. 
This model predicts thermal flux at any wavelength
based on albedo,
diameter, and the object's heliocentric and geocentric
distances. In addition, the model used here includes a
variable beaming parameter, a catch-all factor
accounting for surface roughness and other
irregularities in the object's emission, rather than a
single value to account for this effect for all
objects. The model also allows the millimeter
emmisivity of Trojans as a free parameter to 
be calibrated later.

The model assumes a spherical object
and calculates the instantaneous equilibrium
temperature at each
point on the { day hemisphere}.
For an object receiving a solar flux $S$ with bond
albedo $A$, bolometric emissivity $\epsilon$, and
beaming parameter $\mu$, the temperature at each 
angle from the subsolar point $\omega$ is,  with $\sigma$ as the Stefan-Boltzmann constant: 

\begin{equation}\label{eq:thermalmodeltemp}
    T  = \left( \frac{[1-A]S\cos{\omega}}{\mu\epsilon\sigma} \right)^{1/4}.
\end{equation}

The Bond albedo is related to the geometric albedo
$p_v$ as $A = qp_V$, where $q$ is the phase integral.
{ The correct phase integral to be used here is
not obvious, but for the generally low albedos expected here changes in the value of $q$ make only minor changes in derived parameters. We thus use the standard formulation derived by \citet{1989aste.conf..524B} of $q=0.24+0.684G$, where $G$ is the phase parameter. \citet{schemel_zwicky_2020} find a median value of $G=0.24$ for Jupiter Trojans, giving a value of $q=0.45$.}

With the temperatures at each angle from the subsolar
point found, the emission at the wavelength of the ALMA
observations can be calculated and summed over the
surface of the object, accounting for its diameter,
emissivity, and its distance from the Sun and Earth. In
addition, the absolute magnitude $H_V$, albedo, and
diameter $D$ of an object are related as:

\begin{equation}\label{eq:diammagalb}
    H_V = -5\log{_{10}(D\sqrt{p_V}/1330)}
\end{equation}

The { absolute magnitudes used in the WISE analysis} are consistently lower than those found from the ZTF analysis performed in \citet{schemel_zwicky_2020}, with an average 
difference of 0.32 magnitudes for the twelve objects in
our sample. 
Applying this offset to the full WISE dataset gives a new
mean Jupiter Trojan albedo of 0.051.
We take this as the canonical value and now 
use the three calibrator targets 
(18054, 18137, and 32501) to determine an average
value for millimeter emissivity.
To do so we 
create a likelihood model relating our
model parameters to the observations by using the 
thermal model to produce a flux density
and an absolute magnitude given diameter, albedo,
and beaming parameter as input parameters. 
We use this likelihood model in an MCMC model
to sample this phase space,
again employing the {\it emcee} package
from \citet{foreman-mackey_emcee_2013}. 
We have no observational constraints on beaming
parameter in our data, so we use the full 
distribution of values found in
\citet{2012ApJ...759...49G} as our prior. Priors in
the other parameters were uniform and positive
(and less than one, for albedo). 
All MCMC chains were run to a length of 5000 samples,
more than
100 times the autocorrelation length, and the first
500 samples of each chain were discarded as burn-in.
The marginalized posterior distributions are
nearly gaussian, and with the expected anti-correlation
between diameter and albedo. 
We manually adjust the emissivity of the calibrator
objects to force the three objects to have an average 
albedo of 0.051. Our derived emissivity is 0.753.

Using this derived emissivity for our entire sample,
we now run the MCMC model on all nine putatively high-albedo
targets, again running each chain more than 100 times
the autocorrelation length.
We report 
median values of the derived parameters
with uncertainties giving the 14\% and
86\% percentiles of the distribution in Table \ref{tab:diamalbedo}.

\begin{deluxetable*}{lcccc}
\label{tab:diamalbedo}
\tablecaption{Diameter and albedos obtained using the MCMC procedure to determine best fit thermal models. Errors were obtained using the distribution of each parameter from the MCMC procedure. If the distribution was relatively symmetric, only a single uncertainty is given; otherwise, upper and lower errors are both stated. If the object has multiple possible rotational periods, the average albedo with the maximum range of uncertainty is also given on the first entry for each object.}
\tablehead{
\colhead{object} & \colhead{period (h)} & \colhead{diameter (km)} & \colhead{albedo} & \colhead{adopted albedo}}
\startdata
05123    & $9.897 \pm 0.008$  & $50.3_{-1.1}^{+1.2}$  & $0.063_{-0.005}^{+0.006}$ & $0.063_{-0.005}^{+0.006}$ \\ 
08125  & $51.2\pm 0.2$    & $36.4 \pm 0.8$   & $0.052 \pm 0.004$ & $0.052 \pm 0.004$ \\
11488  & ||                 & $22.7 \pm 1.4$ & $0.09\pm0.01$ & $0.091_{-0.010}^{+0.012}$ \\
13331   & $373.5 \pm 13.8$ & $17.7 \pm 0.9$ & $0.132_{-0.013}^{+0.015}$ & $0.132_{-0.013}^{+0.015}$ \\   
13372    & ||             & $25.4_ \pm 0.9$ & $0.080 \pm 0.006$ & $0.080 \pm 0.006$ \\
13694    & $19.95 \pm 0.02$  & $30.4_{-0.8}^{+0.9}$ & $0.068 \pm 0.005$ &  $0.069_{-0.005}^{+0.006}$ \\
             & $34.14 \pm 0.06$  & $30.4 \pm 0.9 $ & $0.070 \pm 0.005 $& \\
18054 \tablenotemark{a}   & ||    & $37.4 \pm 1.5$ & $0.063_{-0.005}^{+0.006}$&$0.063_{-0.005}^{+0.006}$ \\
18137 \tablenotemark{a}  & $12.102 \pm 0.007$ & $33.6 \pm 1.6 $ & $0.049 \pm 0.005$ &   $0.050_{-0.007}^{+0.009}$  \\ 
                                    & $16.20 \pm 0.01$  & $33.7_{-1.6}^{+1.5}$ & $0.049_{-0.005}^{+0.006}$ & \\ 
                                    & $24.47 \pm 0.05$  & $33.7_{-1.6}^{+1.5}$ & $0.053_{-0.005}^{+0.006}$ & \\
18263   & $10.330\pm 0.006$  & $20.1 \pm 0.7$ & $0.102_{-0.008}^{+0.009}$ & $0.102_{-0.008}^{+0.009}$\\
24452  & ||                 & $21.9_{-1.6}^{+1.5}$ & $0.06 \pm 0.01$ & $0.06 \pm 0.01$ \\ 
32501 \tablenotemark{a}     & ||& $35.5_{-2.1}^{+2.0}$ & $0.039_{-0.004}^{+0.005}$ & $0.039_{-0.004}^{+0.005}$ \\ 
42168   & $4.499\pm 0.002$   & $18.1 \pm 1.1$ & $0.13 \pm 0.02$ &  $0.13_{-0.03}^{+0.04}$ \\
                   & $4.966\pm 0.002$   &  $18.1 \pm 1.1$ & $0.14 \pm 0.02$ &  \\
                   & $5.540\pm 0.002$   & $18.1 \pm 1.1$ & $0.15 \pm 0.02$& \\
                   & $6.265\pm 0.003$   & $18.1 \pm 1.1$ &  $0.15 \pm 0.02$ &\\
                   & $7.206\pm 0.004$   & $18.1 \pm 1.1$ & $ 0.13\pm0.02$  &\\
                   & $8.482\pm 0.005$   & $18.1 \pm 1.1$ & $0.118_{-0.014}^{+0.017}$ & \\
                   & $10.308\pm 0.008$  & $18.1 \pm 1.1$ & $0.115_{-0.013}^{+0.016}$ & \\
                   & $13.14\pm 0.01$    & $18.1\pm 1.1$ & $0.130_{-0.015}^{+0.019}$ &\\
\enddata
\tablenotetext{a}{emissivity calibrator}
\end{deluxetable*}

\begin{figure}
    \centering
    \includegraphics[width=8cm]{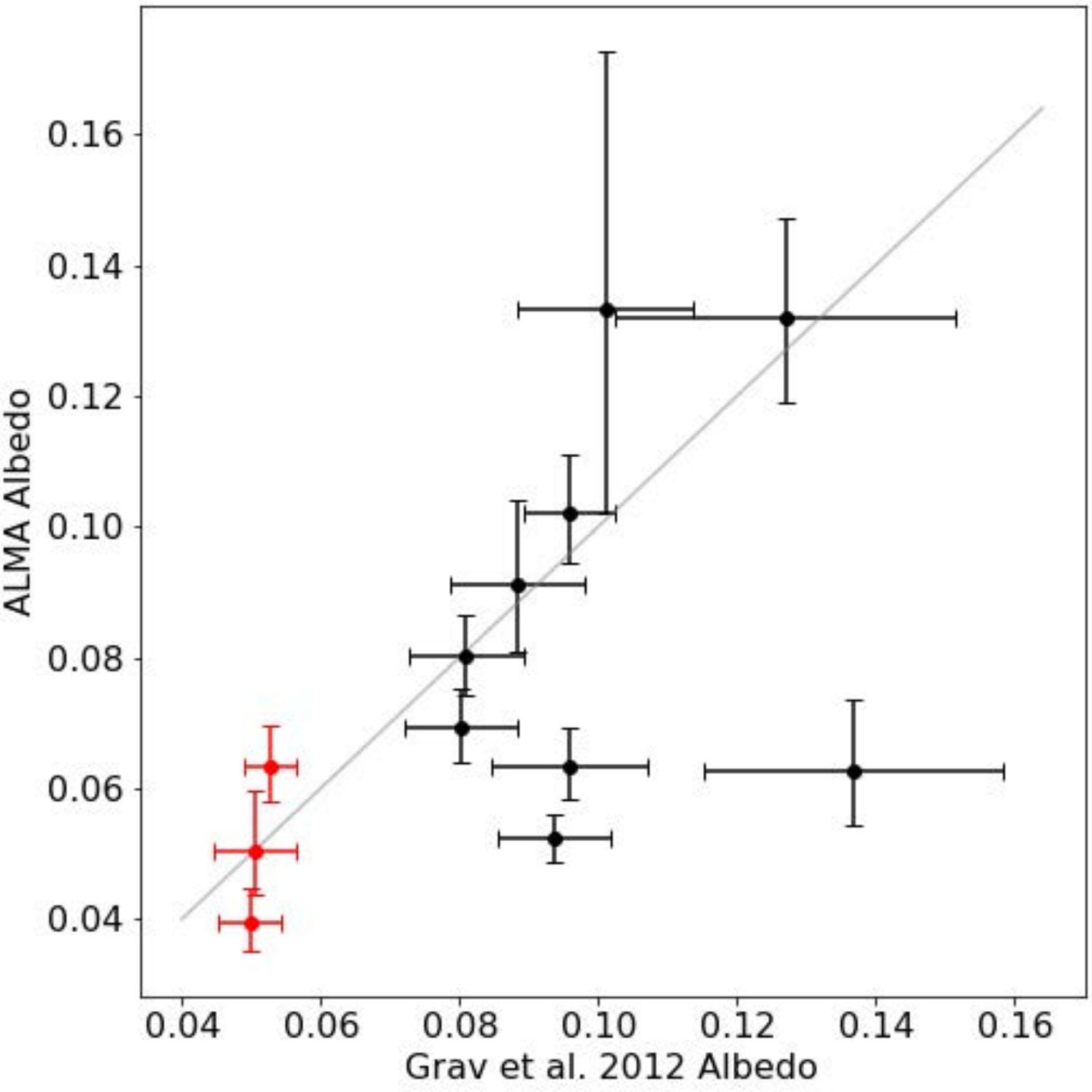}
    \caption{Comparison of albedos obtained using ALMA data to the rescaled albedos of \citet{2012ApJ...759...49G}. The grey line is of slope 1, so that if the same result was found for an object in both analyses, it would fall on that line. The
    three objects used for emissivity calibration are shown in red.}
    \label{fig:almavwise}
\end{figure}

\begin{figure}
    \centering
    \includegraphics[width=8cm]{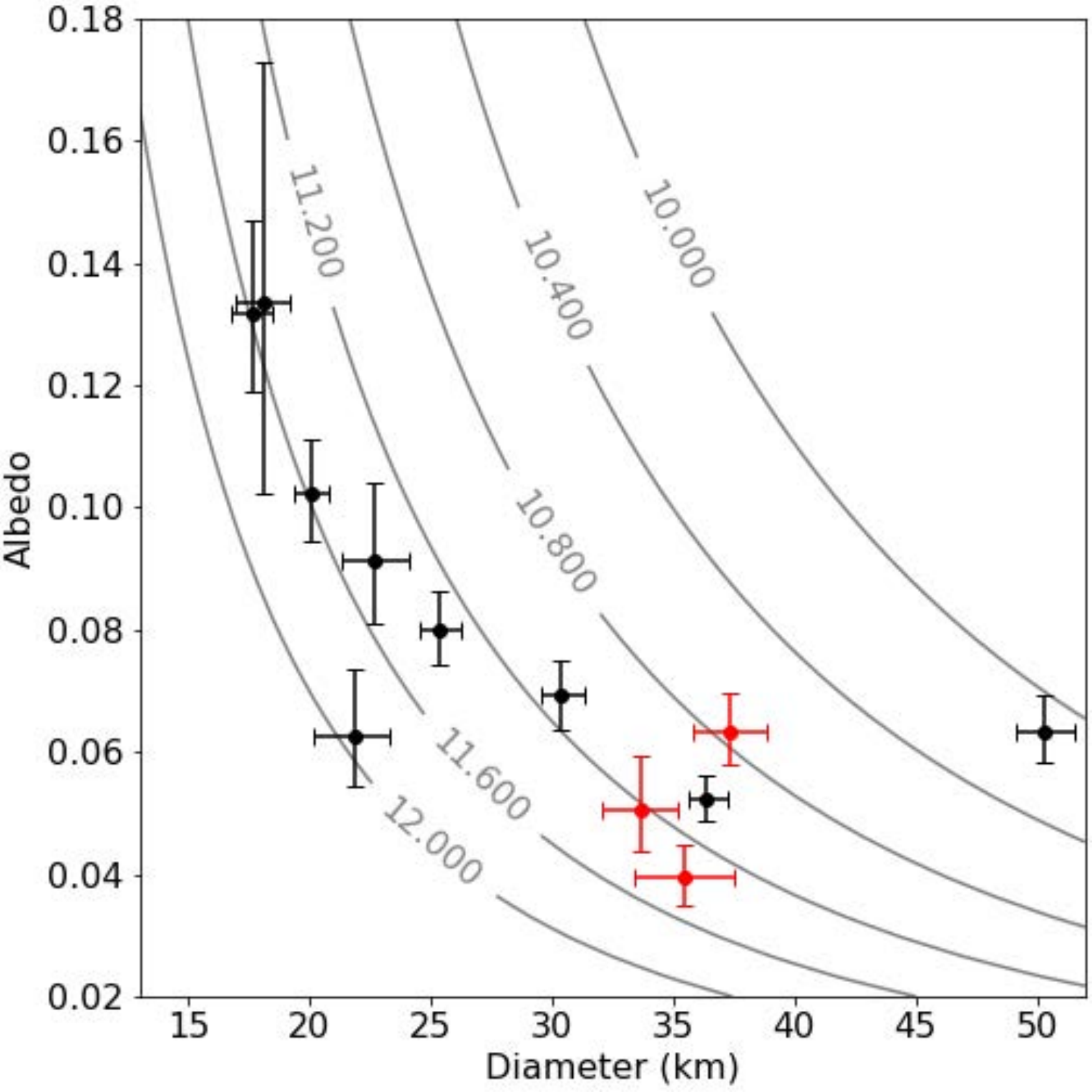}
    \caption{Diameter plotted against albedo for all objects modeled, with the three calibrators shown in red. Lines of constant absolute magnitude (Equation \ref{eq:diammagalb}) are plotted in gray for values of absolute magnitude between 10 and 12. While it appears
    that diameter and albedo are anti-correlated, this 
    apparent trend is driven mostly by the fact that for a given value of absolute magnitude, albedo and diameter
    can only fall along the indicated line. 
 Because of the limited range of absolute magnitudes observed,
 it is unknown if
    small low albedo objects exists, though from the WISE
    data it is clear that 
    large high albedo objects do not.}
    \label{fig:diamvalb}
\end{figure}
{ Based on the ZTF data, the spherical 
assumption for these objections is clearly
incorrect. We perform simple experiments
to determine if the derived albedos are 
systematically affected by this assumption. 
We create a series of ellipsoids with fixed
albedos of 0.05, with random
dimensions, observed at random angles, and
with varying photometric functions, and we
calculate the thermal flux density and
absolute magnitude that would be measured
for each object. We then insert these measurements
into our thermal model and determine the
albedo that would be inferred. We find 
an standard deviation of 6\% between the 
derived values and the simulated values, and
no systematic offset. We thus conclude that
unknown elongation and viewing angle effects
contribute an additional $\sim$6\% uncertainty.}

\section{Discussion} \label{sec:discussion}
Of the nine 
Jupiter Trojans for which WISE derived
unusually high albedos, we find that
four have ALMA-derived albedos less than 0.07,
consistent with typical values
in the larger population (Fig. \ref{fig:almavwise}).
We have no reason to believe that these
objects have changed since the time of the WISE
observations, { and the discrepancies between WISE assumed absolute magnitudes 
and those used here are uncorrelated
with the magnitude (or sign) of the discrepancy between the WISE-derived albedos and those derived here.} The higher S/N of the ALMA
observations and the lower sensitivity to model
assumptions suggests that the ALMA albedos
are more reliable and that for some of these
small objects WISE does indeed suffer from
random and systematic uncertainties larger
than the formal error bars.

Of the five remaining targets, four have ALMA-derived
albedos consistent with those from WISE, while one has
an even larger albedo than reported by WISE.
With these observations completely independent 
and no reason to otherwise expect these objects
to have high albedos,
we regard the match between
the ALMA and WISE albedos as strong confirmation
of the higher-than-typical albedos of these objects.
The
two smallest objects, 13331 and 42168, have albedos
nearly double that of the main Trojan population.

In our sample, albedos decrease systematically at smaller sizes (Fig. \ref{fig:diamvalb}). Some of the decrease
is a simple observational bias: objects with the
same absolute magnitude but smaller sizes
must necessarily have higher albedos.
The absolute magnitudes of the majority of
our
sample are all between 10.8 and 11.8. 
The diameter vs. size of these objects must
{ lie} on a curve like shown in Figure \ref{fig:diamvalb}.
Small low albedo objects, if they exist, 
would have low absolute magnitudes and would
not be included in our sample. Examination of
the WISE results in Figure 1 suggests that 
such objects do indeed exist, though, again,
unaccounted-for uncertainties could dominate
at these small sizes in the WISE data. In
contrast, large high albedo Trojans are
unlikely to exist, as they would have been
detected with high S/N in the WISE data.

Overall the ALMA results support the conclusion
that a modest number of 15-25 km diameter
Jupiter Trojans have albedos elevated above
those of the general population. 
The sizes and numbers of these objects are
consistent with the hypothesis that they are
the expected larger diameter tail of 
recent collisional fragments. If this hypothesis
is correct, we would expect a larger fraction
of high albedo Trojans at smaller sizes.
In addition, we would expect that these
high albedo mid-sized Trojans could be
the most likely location to find spectroscopic
hints of the interior composition of
Jupiter Trojans. Continued study of this
population could yield critical insight into
the formation location of these objects, and,
by extension, into the dynamical history
of the solar system.
\acknowledgements
This research is supported by the Caltech Student-Faculty Programs WAVE Fellowship funded by the Carl F. Braun Trust and by  NSF Astronomy \& Astrophysics Research Grant \#2109212. 
This paper makes use of the following ALMA data: ADS/JAO.ALMA\#2019.1.01158.S. ALMA is a partnership of ESO (representing its member states), NSF (USA) and NINS (Japan), together with NRC (Canada), MOST and ASIAA (Taiwan), and KASI (Republic of Korea), in cooperation with the Republic of Chile. The Joint ALMA Observatory is operated by ESO, AUI/NRAO and NAOJ.The National Radio Astronomy Observatory is a facility of the National Science Foundation operated under cooperative agreement by Associated Universities, Inc.
Based on observations obtained with the Samuel Oschin
Telescope 48 inch Telescope at the Palomar Observatory as
part of the Zwicky Transient Facility project. Major funding
has been provided by the U.S. National Science Foundation
under grant No. AST-1440341 and by the ZTF partner
institutions: the California Institute of Technology, the Oskar
Klein Centre, the Weizmann Institute of Science, the
University of Maryland, the University of Washington,
Deutsches Elektronen-Synchrotron, the University of Wisconsin-
Milwaukee, and the TANGO Program of the University
System of Taiwan.
This work uses data obtained from the Asteroid Lightcurve Data Exchange Format (ALCDEF) database,
which is supported by funding from NASA grant 80NSSC18K0851.

\appendix
\section{Asteroid Periodograms and Phased Lightcurves}
The following figures reproduce Figure \ref{fig:lightcurves} for all objects for which periods were obtained.

\begin{figure}[H]
    \centering
    \includegraphics[width=17cm]{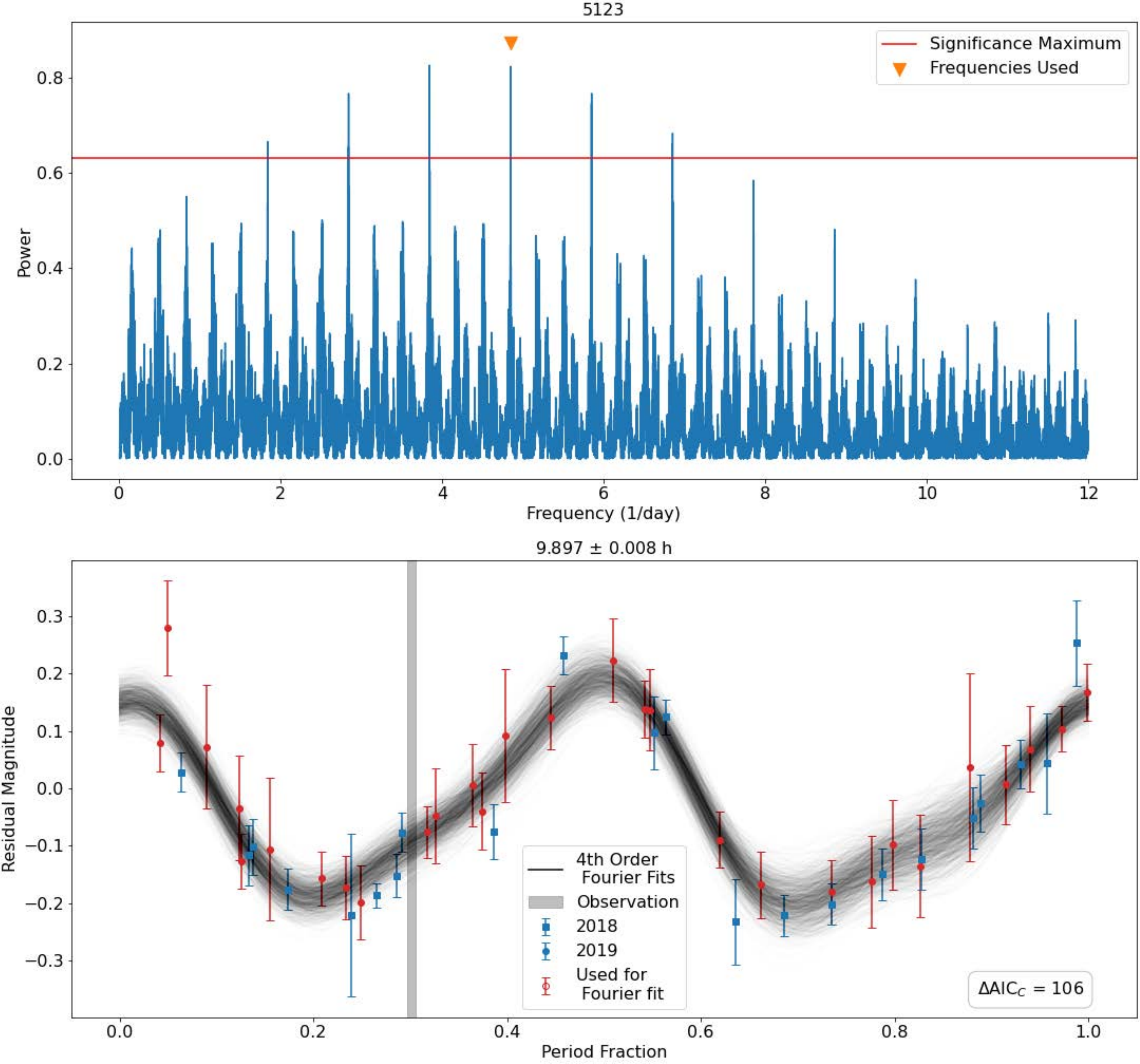}
    \label{fig:5123}
\end{figure}

\begin{figure}[H]
    \centering
    \includegraphics[width=17cm]{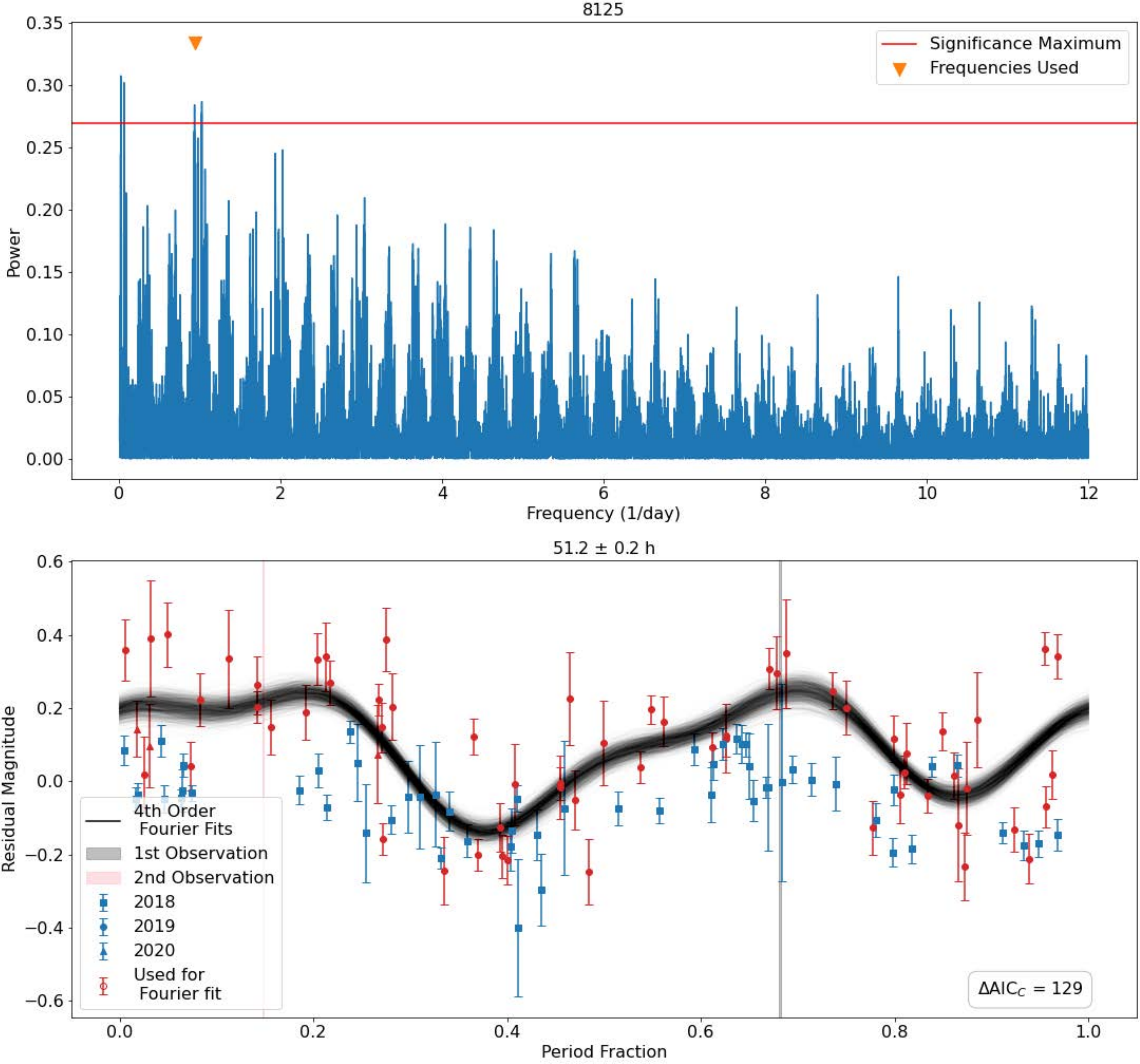}
    \label{fig:8125}
\end{figure}

\begin{figure}[H]
    \centering
    \includegraphics[width=17cm]{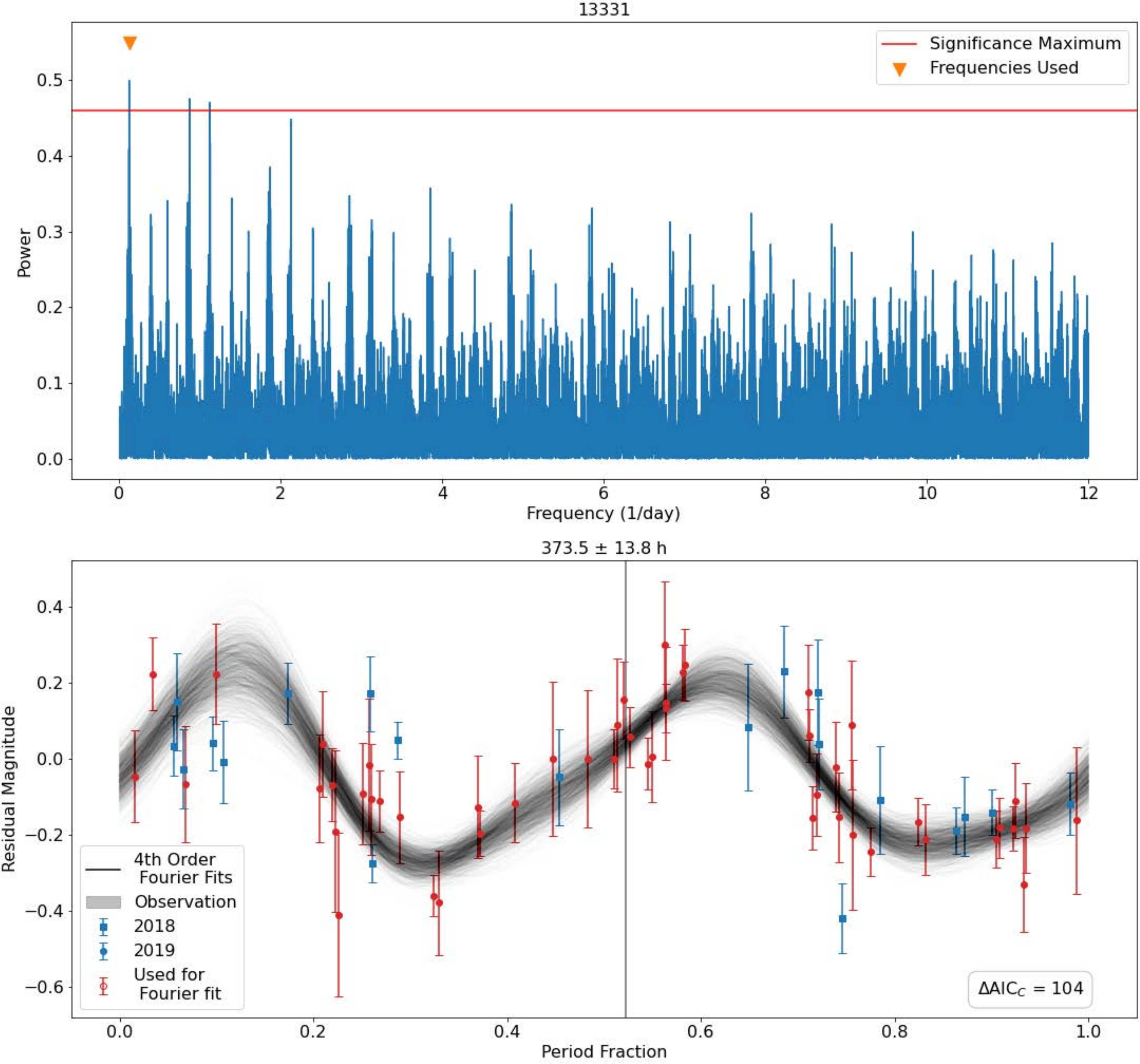}
    \label{fig:13331}
\end{figure}

\begin{figure}[H]
    \centering
    \includegraphics[width=17cm]{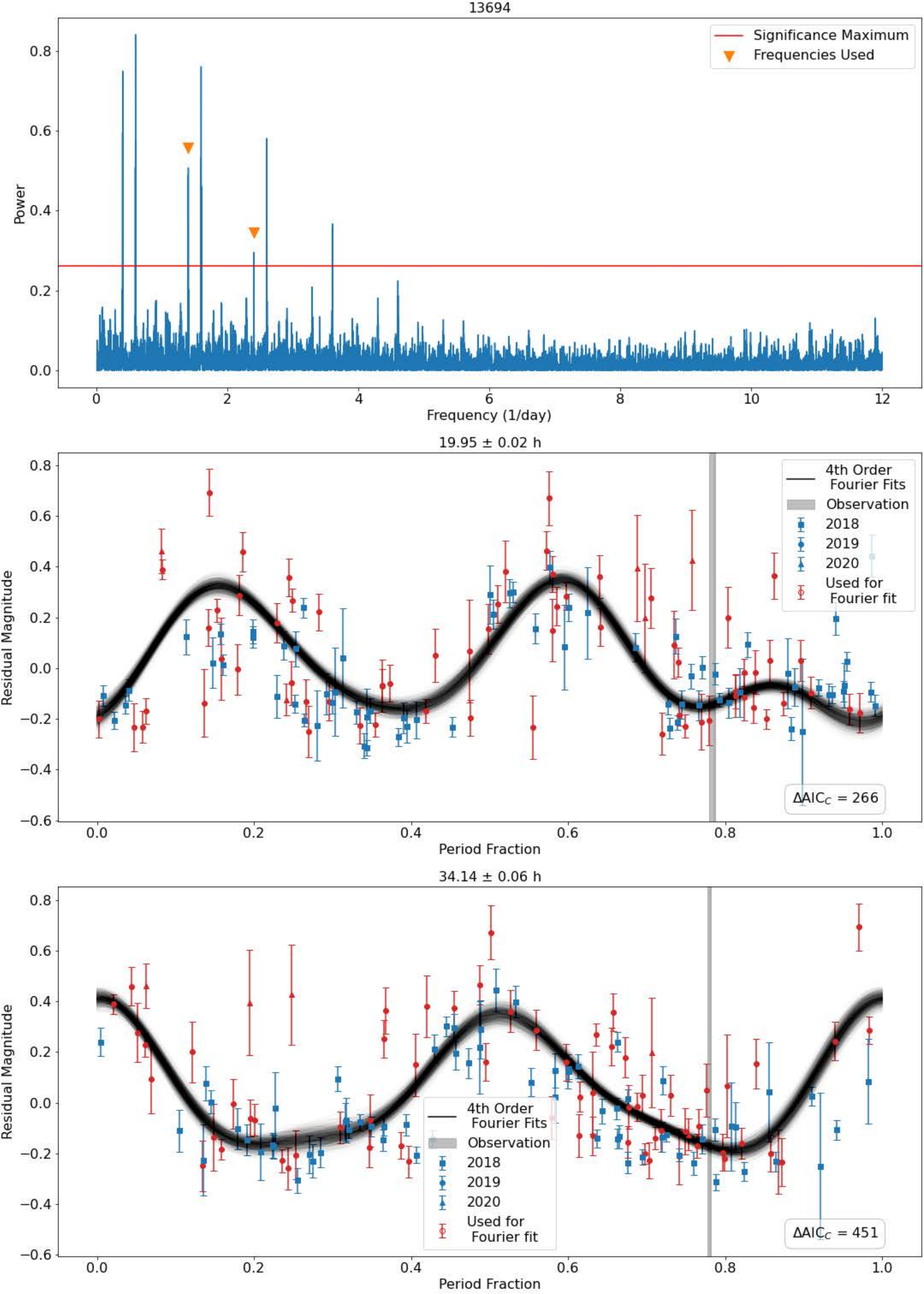}
    \label{fig:13694}
\end{figure}

\begin{figure}[H]
    \centering
    \includegraphics[width=13cm]{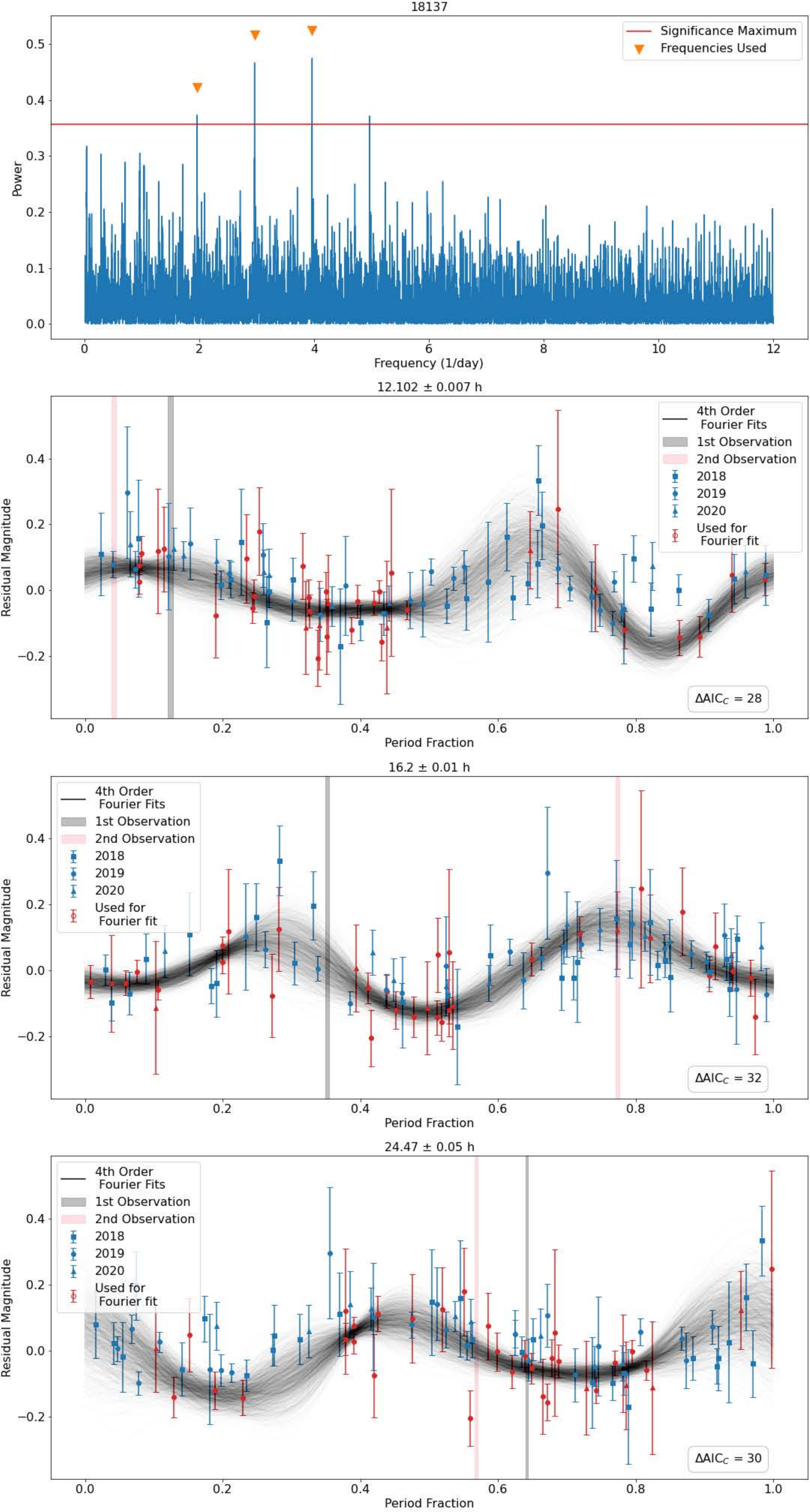}
    \label{fig:18137}
\end{figure}

\begin{figure}[H]
    \centering
    \includegraphics[width=17cm]{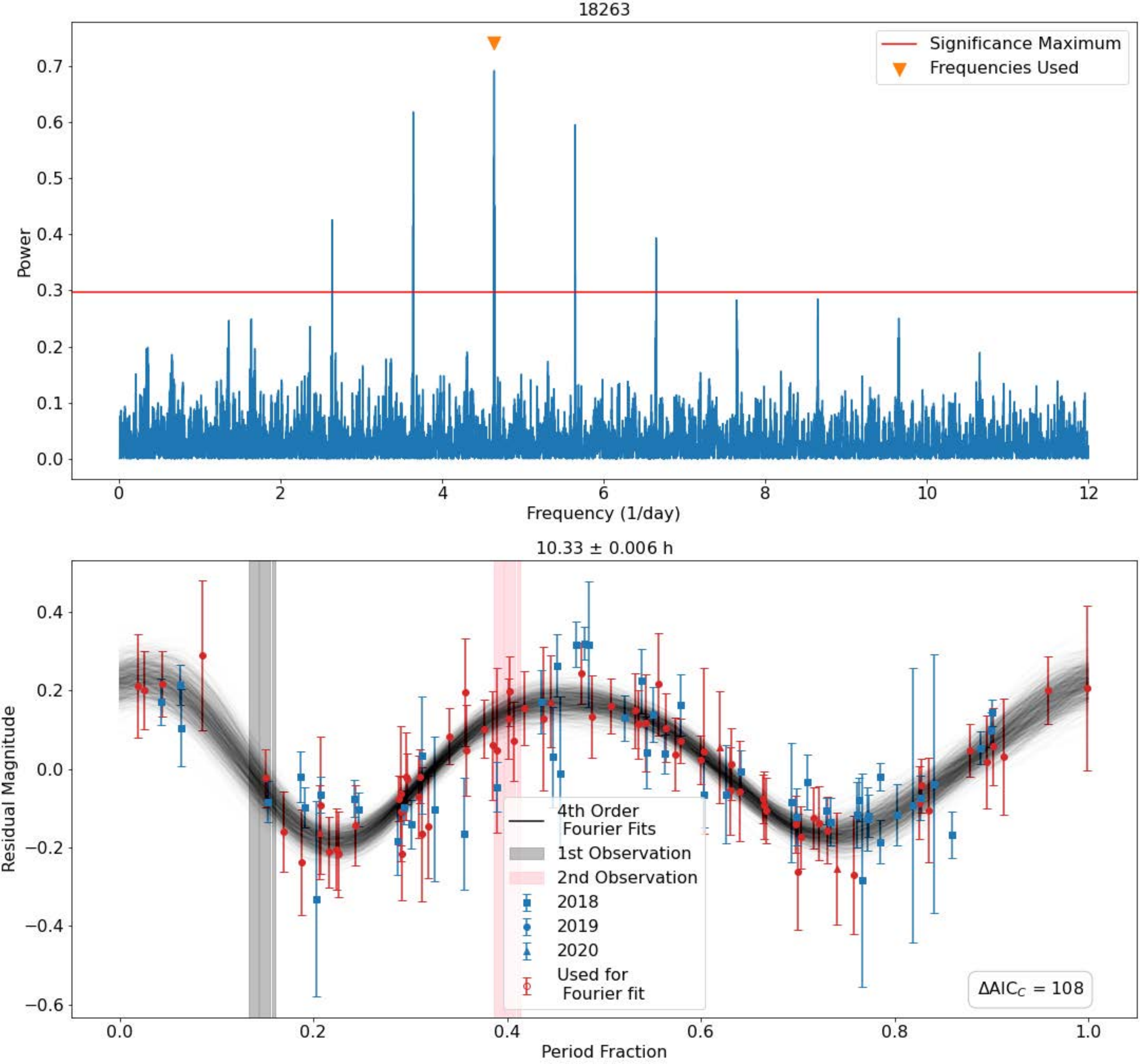}
    \label{fig:18263}
\end{figure}

\begin{figure}[H]
    \centering
    \includegraphics[width=13cm]{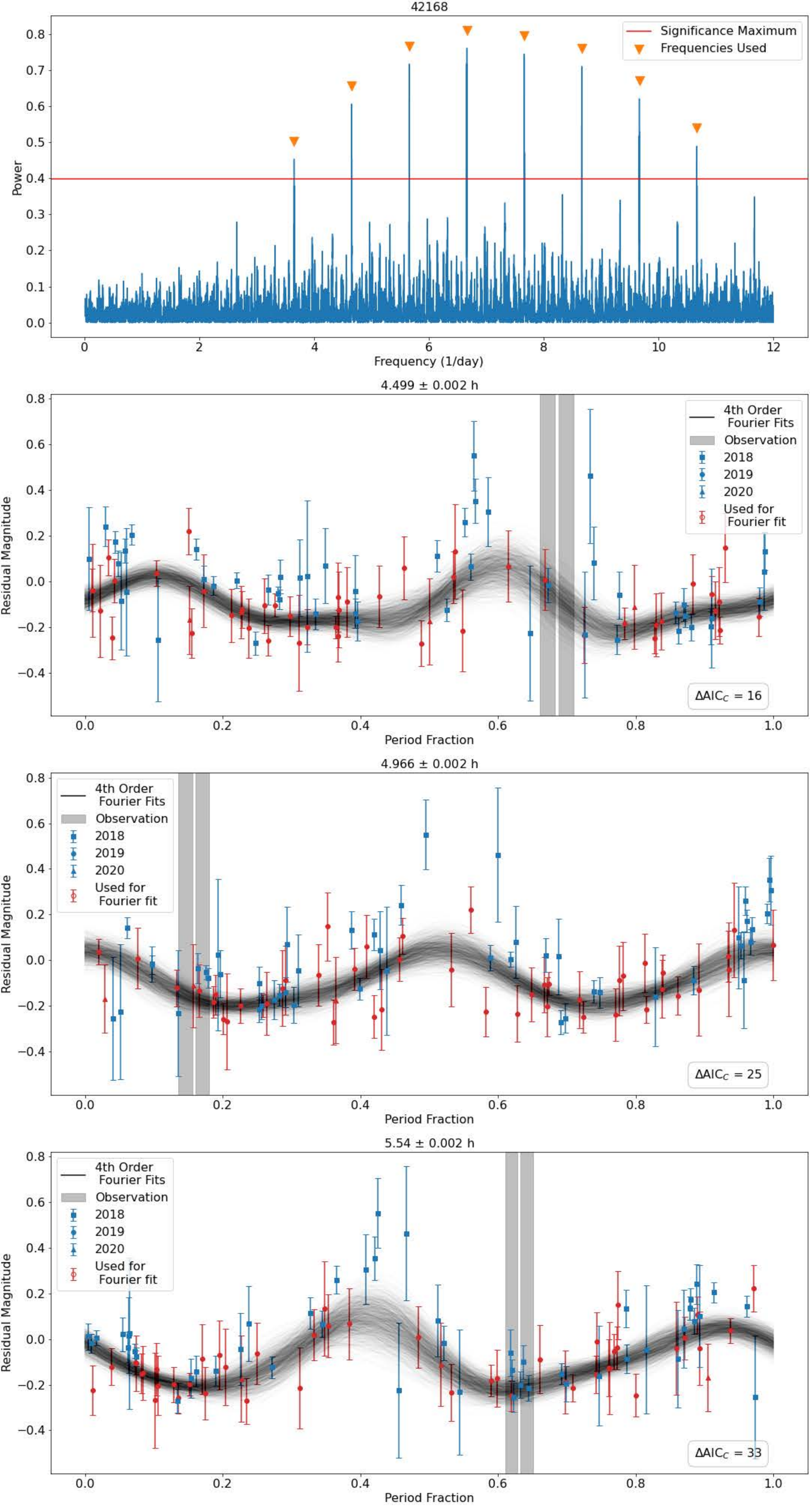}
    \label{fig:42168_TOP}
\end{figure}
\begin{figure}[H]
    \centering
    \includegraphics[width=13cm]{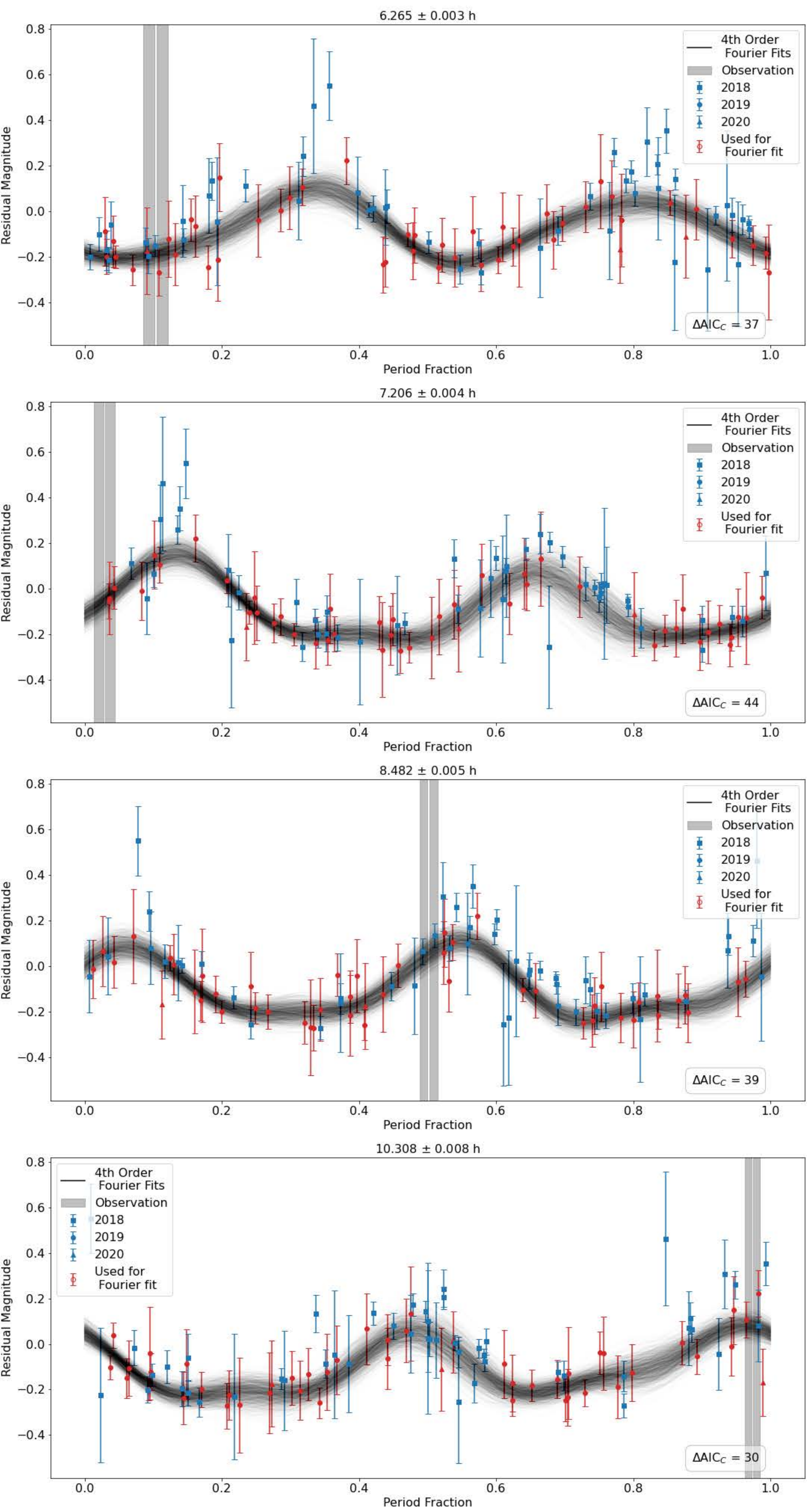}
    \label{fig:42168_MID}
\end{figure}
\begin{figure}[H]
    \centering
    \includegraphics[width=13cm]{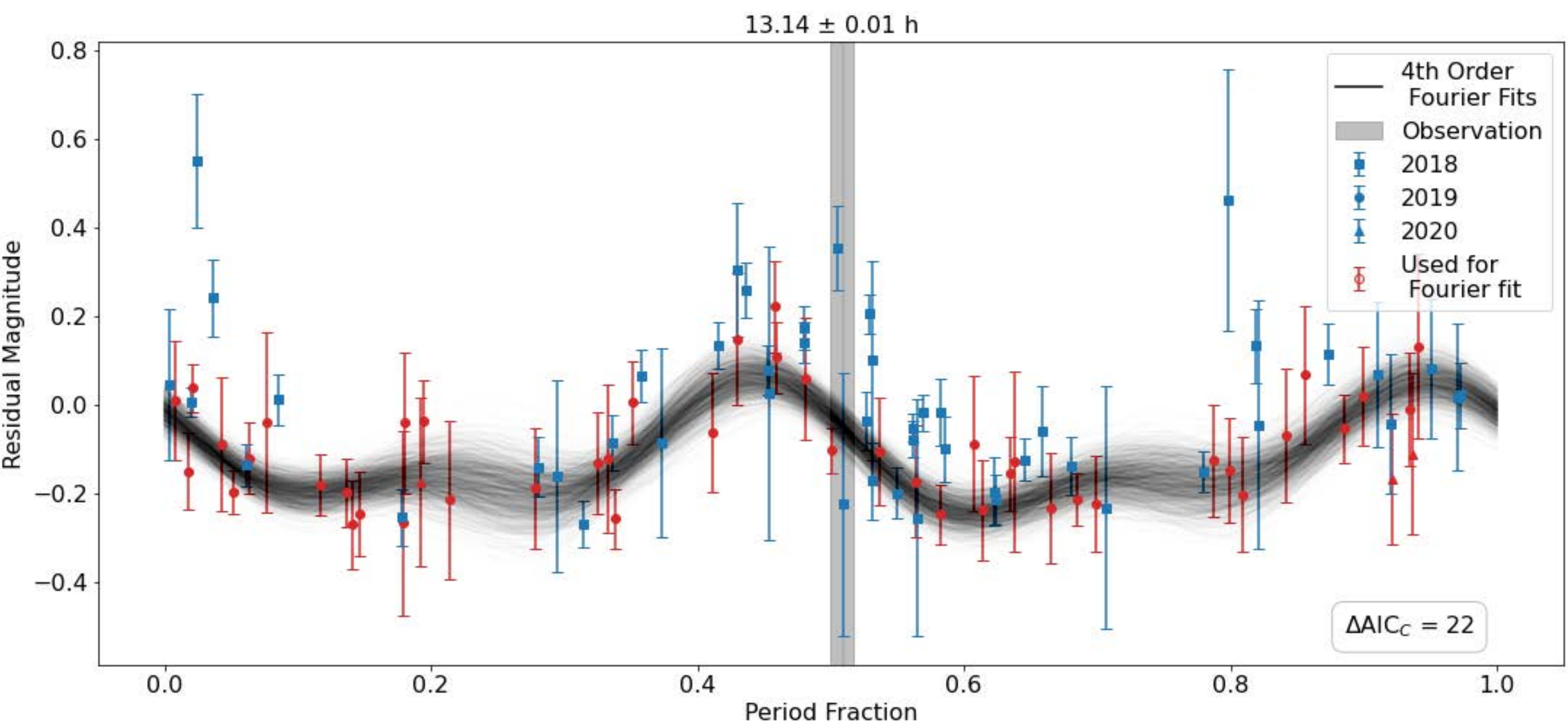}
    \label{fig:42168_BOTTOM}
\end{figure}

\end{document}